# Statefinder analysis of universe models with a viscous cosmic fluid and a fluid with non-linear equation of state

Øyvind Grøn

Oslo University College, Department of Engineering, St. Olavs Pl. 4, 0136 Oslo, Norway

Institute of Physics, University of Oslo, P. P. Box 1048 Blindern, 0316 Oslo, Norway

**Abstract**

In the present article we analyze, by means of the statefinder parameter formalism, some universe models introduced by Brevik and co-workers. We determine constants that earlier were left unspecified, in terms of observable quantities. It is verified that a Big Bang universe model with a fluid having a certain non-linear equation of state behaves in the same way as a model with a viscous fluid.

## 1. Introduction

The statefinder parameters were introduced into cosmology by Alam and co-workers[1] in 2003. They were then applied to flat universe models with dark energy and cold matter, where the dark energy was either of the quintessence type or a Chaplygin gas. The latter models were further investigated by Gorini and co-workers[2]. The formalism was generalized to curved universe models by Evans et al.[3].

The formalism was later applied to several universe models with other properties such as interaction between energy and matter[4]-[8] or other types of equation of state than that of the quintessence energy, resulting from for example viscosity[9]-[16]. G. Panotopoulos[17] has applied the statefinder diagnosis to brane universe models.

Iver Brevik and co-workers[21]-[31] have investigated the future behavior of several universe models filled with cosmic fluids with different equations of state[32]-[34]. Some of the equations of state may be relevant to fluids with viscosity[35]. They have in particular investigated whether the models arrive at a final so-called Big Rip singularity. M. P. Dabrowski[36] has used statefinders to investigate universe models with future singularities.

In the present work we want to apply the statefinder formalism to some of the universe models considered by Brevik and co-workers.

## 2. Earlier applications of the statefinder formalism

Alam and co-workers[1] first considered flat universe models with cold matter (dust) and dark energy in the form of quintessence obeying the equation of state

$$p_X = w\rho_X \ , \tag{2.1}$$

where $p_X$ is the pressure of the dark energy and $\rho_X$ its density (we use units so that $c=1$). Here $w$ is a function of time. They defined the statefinder parameters $r$ and $s$ as

$$r = \frac{\ddot{a}}{aH^3} \ , \quad s = \frac{r-1}{3(q-1/2)} \ , \tag{2.2}$$

where

$$q = -\frac{\ddot{a}}{aH^2} = -1 - \frac{\dot{H}}{H^2} \tag{2.3}$$

is the deceleration parameter. In terms of the Hubble parameter and its derivatives with respect to cosmic time the statefinder parameters are given by

$$r = 1 + 3\frac{\dot{H}}{H^2} + \frac{\ddot{H}}{H^3} \ , \quad s = -\frac{2}{3H}\frac{3H\dot{H}+\ddot{H}}{3H^2+2\dot{H}} \ . \tag{2.4}$$

The deceleration parameter and the statefinder parameters can also be expressed in terms of the Hubble parameter and its derivative with respect to the redshift, represented by $x \equiv 1+z = 1/a$,

$$q = \frac{H'}{H}x - 1 \quad , \tag{2.5}$$

$$r = 1 - 2\frac{H'}{H}x + \left[\frac{H''}{H} + \left(\frac{H'}{H}\right)^2\right]x^2 \quad , \tag{2.6}$$

$$s = \frac{-2H'x/H + \left[H''/H + (H'/H)^2\right]x^2}{3\left[H'x/H - 3/2\right]} \quad . \tag{2.7}$$

Calculating the deceleration parameter and the statefinder parameters for this class of universe models one finds

$$q = \frac{1}{2}(1+3w) \; , \quad r = 1 + \frac{9}{2}w(1+w)\Omega_X - \frac{3}{2}\frac{\dot{w}}{H}\Omega_X \; , \quad s = 1 + w - \frac{1}{3}\frac{\dot{w}}{wH} \quad . \tag{2.8}$$

Hence

$$r = 1 + \frac{9}{2}w\Omega_X s \tag{2.9}$$

It should be noted that the flat $\Lambda$CDM universe model, which fits all the cosmological observations, has $w = -1$, $\dot{w} = 0$ and hence $(r,s) = (1,0)$.

For the type of dark energy called Chaplygin gas, obeying the equation of state

$$p_C = -A/\rho_C^\alpha \quad , \tag{2.10}$$

where $A$ and α are positive constants, the statefinder parameters are found by using the relationships

$$q = \frac{1}{2}\left(1 + 3\frac{p}{\rho}\right) \; , \quad r = 1 + \frac{9}{2}\frac{\rho + p}{\rho}\frac{\dot{p}}{\dot{\rho}} \; , \quad s = \frac{\rho + p}{p}\frac{\dot{p}}{\dot{\rho}} \quad , \tag{2.11}$$

where $p$ and $\rho$ are the total pressure and density, respectively. Hence,

$$r = 1 + \frac{9}{2}\frac{p}{\rho}s \quad . \tag{2.12}$$

The formalism was generalized to curved universe models by Evans et. Al[3]. Then the state parameter *s* is defined by

$$s = \frac{r - \Omega}{3(q - \Omega/2)} \quad , \tag{2.13}$$

and eqs.(2.11) take the form

$$q = \frac{1}{2}\left(1 + 3\frac{p}{\rho}\right)\Omega \; , \; r = \left(1 - \frac{3}{2}\frac{\dot{p}}{H\rho}\right)\Omega \; , \; s = -\frac{1}{3H}\frac{\dot{p}}{p} \quad . \tag{2.14}$$

Applied to a universe with only a Chaplygin gas this gives[2]

$$r = 1 - \frac{9}{2\alpha} s(1 + s) \quad . \tag{2.15}$$

If the source of the dark energy is a scalar field $\phi$ with the potential $V(\phi)$, the equation of state factor $w$ is

$$w = \frac{\dot{\phi}^2 - 2V(\phi)}{\dot{\phi}^2 + 2V(\phi)} \quad . \tag{2.16}$$

In this case the statefinder parameters are[3]

$$q = \frac{\Omega}{2} + \frac{\kappa}{2H^2}\left(\frac{1}{2}\dot{\phi}^2 - V\right) \; , \; r = \Omega + \frac{3}{2}\kappa\frac{\dot{\phi}^2}{H^2} + \kappa\frac{\dot{V}}{H^3} \; , \; s = 2\frac{\dot{\phi}^2 + \frac{2}{3}\frac{\dot{V}}{H}}{\dot{\phi}^2 - 2V} \quad , \tag{2.17}$$

where $\kappa = 8\pi G$ is Einstein's constant of gravitation.

W. Zimdahl and D. Pavón[4], and X. Zhang[7-9] with co-workers[6] applied the statefinder formalism to universe models with two interacting fluids. The dark matter component *M* interacts with the dark energy described by a scalar field $\phi$ by

$$\dot{\rho}_M + 3H\rho_M = -Q \; , \; \dot{\rho}_\phi + 3H\rho_\phi(1 + w_\phi) = Q \quad . \tag{2.18}$$

The deceleration parameter of such a universe model is

$$q = \frac{1}{2}(1 + 3w_\phi \Omega_\phi) \quad . \tag{2.19}$$

Defining effective equations of state for the dark matter and energy by

$$w_M^{eff} = \frac{Q}{3H\rho_M} \; , \; w_\phi^{eff} = -\frac{Q}{3H\rho_\phi} \quad , \tag{2.20}$$

the statefinder parameters may be written

$$r = 1 - \frac{3}{2}\left[w'_\phi - 3w_\phi\left(1 + w_\phi^{eff}\right)\right]\Omega_\phi \ , \quad s = 1 - \frac{1}{3}\frac{w'_\phi}{w_\phi} + w_\phi^{eff} \ , \tag{2.21}$$

where $w'_\phi$ is $w_\phi$ differentiated with respect to $u = \ln a = -\ln(1+z)$, where $a$ is the scale factor and $z$ the redshift. The relation between $r$ and $s$ is

$$r = 1 + \frac{9}{2}w_\phi s \ . \tag{2.22}$$

M. G. Hu and X. H. Meng[17] have studied flat universe models with a viscous fluid. This type of models shall be investigated in some detail in the next section. They also analyzed a flat universe model with only dark energy obeying the inhomogeneous equation of state,

$$p = w\rho + p_1 \ , \tag{2.23}$$

where $w$ and $p_1$ are both constant. Defining the quantities

$$\tilde{\gamma} = -\frac{p_1}{\rho_0} \ , \quad V = p_1\left(\frac{1}{\rho} - \frac{1}{\rho_0}\right) \ , \tag{2.24}$$

where $\rho_0$ is the present density, they find for the deceleration parameter and the statefinder parameters,

$$q = \frac{3}{2}V - 1 \ , \quad r = 1 + \frac{9}{2}(\tilde{\gamma} - 1)V \ , \quad s = \frac{(\tilde{\gamma} - 1)V}{V - 1} \ . \tag{2.25}$$

The ΛCDM universe model, that is consistent with all present observations and may be considered the standard model of the universe[37], has $(q, r, s) = (-1, 1, 0)$ which is fulfilled for the above model if $\rho = \rho_0$.

**3. Viscous dark energy**

In this section I will deduce the expressions for the statefinder parameters of flat universe models with dust and viscous dark energy in terms of the Hubble parameter. The dark energy is assumed to obey the usual equation of state, $p = w\rho$, with constant value of $w$. Friedmann's 1. equation then takes the form

$$H^2 = (\kappa/3)(\rho_m + \rho_x) = (\kappa/3)\rho_{cr} \ . \tag{3.1}$$

where $\rho_m, \rho_x, \rho_{cr}$ are the density of the matter, the dark energy and the critical density, respectively. Friedmann's 2. equation is

$$\frac{\ddot{a}}{a} + \frac{H^2}{2} = -\frac{\kappa}{2}\bar{p}_x \ , \tag{3.2}$$

where $\kappa = 8\pi G$ and $G$ is Newton's constant of gravitation, and

$$\bar{p}_x = w\rho_x - 3\varsigma H \tag{3.3}$$

is the effective pressure of the dark energy. We shall assume that the viscosity coefficient $\varsigma$ is constant, and that the dust and the dark energy does not interact. Inserting eq.(3.3) into eq.(3.2) we obtain,

$$\frac{\ddot{a}}{a} = -\frac{H^2}{2} - \frac{\kappa}{2}w\rho_x + \frac{3}{2}H\kappa\varsigma \ . \tag{3.4}$$

By means of eq.(3.4) and introducing the mass parameters

$$\Omega_x = \frac{\kappa}{3H^2}\rho_x \ , \ \Omega_m = \frac{\kappa}{3H^2}\rho_m = 1 - \Omega_x \ , \tag{3.5}$$

where we have used that the universe is flat, the expression (2.3) of the deceleration parameter can be written

$$q = \frac{1}{2}\left(1 + 3w\Omega_x - 3\kappa\varsigma/H\right) \ . \tag{3.6}$$

From eqs.(2.3) and (2.4) we get

$$r = q + 2q^2 - \dot{q}/H \ . \tag{3.7}$$

The equation of continuity of the dark matter is

$$\dot{\rho}_m = -3H\rho_m \ , \tag{3.8}$$

and of the dark energy

$$\dot{\rho}_x = -3H\left[(1+w)\rho_x - 3H\varsigma\right] \ . \tag{3.9}$$

Differentiating eq.(3.1) and inserting eqs. (3.8) and (3.9) we obtain

$$\frac{\dot{H}}{H^2} = -\frac{3}{2}\left(1 + w\Omega_x - \frac{\kappa\varsigma}{H}\right) \ . \tag{3.10}$$

Differentiating the first of eqs.(3.5) and using eqs.(3.9) and (3.10) we find

$$\dot{\Omega}_x = 3(1-\Omega_x)(\kappa\varsigma - w\Omega_x H) \ . \tag{3.11}$$

Differentiating the expression (3.6) and using eqs.(3.10) and (3.11) then gives

$$\dot{q} = -\frac{9}{2}(1-\Omega_x)\Omega_x w^2 H - \frac{9}{4}\left(1 - 2w + 3w\Omega_x - \frac{\kappa\varsigma}{H}\right)\kappa\varsigma \ . \tag{3.12}$$

Inserting the expressions (3.6) and (3.12) into eq.(3.7) finally gives

$$r = 1 + \frac{9}{2}w(1+w)\Omega_x - \frac{9}{4}\left(1+2w+w\Omega_x - \frac{\kappa\varsigma}{H}\right)\frac{\kappa\varsigma}{H} .$$  (3.13)

and

$$s = \frac{1}{2}\left(1+2w - \frac{\kappa\varsigma}{H} + \frac{1}{1-\kappa\varsigma/w\Omega_x H}\right)$$  (3.14)

**4. Dark fluid with a non-linear equation of state**

In ref. 21 Brevik and co-workers have studied, in their second case, a flat Friedmann-Robertson-Walker universe model dominated by a dark energy with equation of state

$$p = w\rho + A\sqrt{\rho} ,$$  (4.1)

where $w$ and $A$ are constants. Lorentz invariant dark energy, which may be represented by a cosmological constant, has $w = -1$ and $A = 0$. For this model Friedmann's 1. equation reduces to

$$H^2 = \frac{\kappa}{3}\rho ,$$  (4.2)

and the equation of continuity takes the form

$$\dot{\rho} + \sqrt{3\kappa}(1+w)\rho^{3/2} + \sqrt{3\kappa}A\rho = 0 .$$  (4.3)

Integration of eq.(4.3) with $\rho(0) = \left[A/(1+w)\right]^2$ gives

$$\rho = \frac{A^2}{\left[e^{\frac{\sqrt{3\kappa}}{2}A\bar{t}} - (1+w)\right]^2} ,$$  (4.4)

where the time co-ordinate used by Brevik and co-workers[21] has been called $\bar{t}$. Note that $\rho(\bar{t}_S) = \infty$ for $\bar{t}_S = -t_A \ln(1+w)$ where

$$t_A = -\left(2/\sqrt{3\kappa}A\right) .$$  (4.5)

The minus is included because a later comparison with observational data will show that $A < 0$. Note also that $\bar{t}_S > 0$ for $w < 0$. I will interpret the singularity as the Big Bang of this universe model. With this interpretation the points of time $\bar{t} < \bar{t}_S$ are without physical significance. Introducing a cosmic time $t = \bar{t} - \bar{t}_S$ with origin at the Big Bang, the expression of the density of the cosmic fluid becomes

$$\rho = \left(\frac{A}{1+w}\right)^2 \frac{1}{\left(1-e^{-t/t_A}\right)^2} \quad . \tag{4.6}$$

From eqs.(4.4), (4.5) and (4.6) follow that the Hubble parameter is

$$H = \frac{H_A}{1-e^{-t/t_A}} \quad , \quad H_A = \frac{2}{3(1+w)t_A} \quad . \tag{4.7}$$

For $t \gg t_A$ the behavior of the expansion approaches that of DeSitter space with a constant Hubble parameter $H_A$ containing a cosmic fluid with density $\rho_A = \left[A/(1+w)\right]^2$. At the initial singularity the Hubble parameter is infinitely great. Normalizing the scale parameter so that it has the value $a(t_0) = 1$ at the present time $t_0$ we find

$$a = \left(\frac{e^{t/t_A}-1}{e^{t_0/t_A}-1}\right)^{\frac{2}{3(1+w)}} \quad . \tag{4.8}$$

Using that

$$H_A t_A = \frac{2}{3(1+w)} \quad , \tag{4.9}$$

the deceleration parameter is

$$q = -1 + \frac{3}{2}(1+w)e^{-t/t_A} \quad . \tag{4.10}$$

Note that for $t \ll t_A$ the scale factor is to lowest order in $t/t_A$

$$a(t) \approx \left(\frac{t}{t_A}\right)^{\frac{2}{3(1+w)}} \quad . \tag{4.11}$$

which is the same behaviour as that of a universe dominated by a perfect fluid obeying the homogeneous equation of state $p = w\rho$. The late time behavior is that of a de Sitter universe with accelerated expansion independently of the value of $w$. For $w > -1/3$ there is a transition from decelerated to accelerated expansion at the point of time

$$t_2 = t_A \ln\left[\frac{3}{2}(1+w)\right] \quad . \tag{4.12}$$

This behavior reflects the fact that the first term of eq.(4.1) dominates initially when the density is large, but the late time behavior is dominated by the second term in eq.(4.1).

Using eqs.(2.4) we find that the statefinder parameters for this universe model are

$$r = 1 - \frac{9}{4}\left(1-w^2\right)e^{-t/t_A} + \frac{9}{4}\left(1+w\right)^2 e^{-2t/t_A} \ , \tag{4.13}$$

$$s = \frac{1}{2}(1+w)\frac{1-w-(1+w)e^{-t/t_A}}{e^{t/t_A}-1-w} \ . \tag{4.14}$$

These expressions are plotted in a $(r,s)$-diagram in Figure 1.

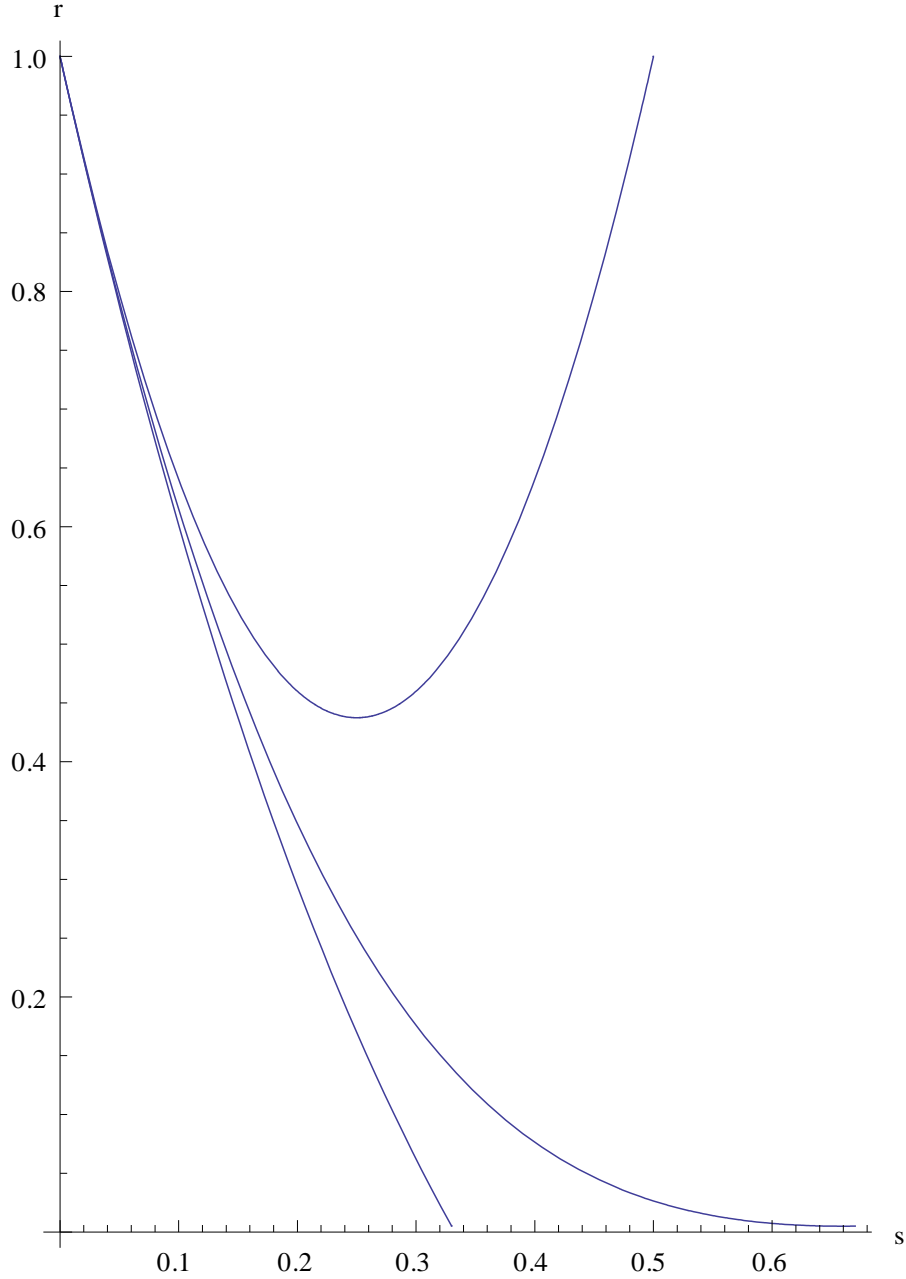

**Figure 1.** $(r,s)$-diagram for the universe model with statefinder parameters given in eqs.(4.13) and (4.14). The curves are plotted for $t \in [0,\infty]$. The lowest curve is for $w = -2/3$, the next one for $w = -1/3$, and the upper one for $w = 0$.

The present values of the Hubble parameter, $H_0 = H(t_0)$, and the deceleration parameter, $q_0 = q(t_0)$ are determined by observations, which show that $H_0 t_0 \approx 1$ and $q_0 \approx -0.6$. This will be used to determine the quantities $H_1$ and $t_A$. From eqs.(4.7) and (4.10) we find that $H_A$ is determined by the equation

$$\frac{H_0}{H_A} + \frac{1+q_0}{\ln\left(1+\dfrac{H_A}{H_0}\right)} = 1 \quad . \tag{4.15}$$

Inserting $q_0 \approx -0.6$ and solving the equation numerically, we find $H_A = 0,8 H_0$. Hence,

$$t_A = -\frac{t_0}{\ln\left(1-\dfrac{H_A}{H_0}\right)} \approx 0.6 t_0 \quad . \tag{4.16}$$

With $t_0 = 13,7 \cdot 10^9$ years we get $t_A \approx 8.2 \cdot 10^9$ years, so that $H_A t_A \approx 0.8 H_0 \cdot 0.6 t_0 \approx 0.5$. According to eq.(4.9) the equation of state factor $w$ then is

$$w = \frac{2}{3 H_A t_A} - 1 \approx \frac{1}{3} \quad . \tag{4.17}$$

This universe model thus contains a fluid behaving somewhat like a combination of electromagnetic radiation and dark energy with an equation of state with negative pressure (note that $A < 0$).

The value $s = 0$ of the $\Lambda CDM$ - model takes place at the point of time

$$t_3 = t_A \ln \frac{1+w}{1-w} \quad . \tag{4.20}$$

A positive value of $t_3$ requires $w > 0$, which is compatible with the value of $w$ obtained from the present values of the Hubble parameter and the deceleration parameter.

**5. Viscous cosmic fluid**

I. Brevik and O. Gorbunova[24] have recently investigated some universe models dominated by a viscous cosmic fluid. They were particularly interested in the late time behaviour of the models and whether they would enter a so-called Big Rip. Therefore they investigated models with $w < -1$. The effective pressure is given by eq.(3.3). I will consider universe models dominated by a viscous fluid with $w > -1$. Defining

$$H_\varsigma = \frac{\kappa \varsigma}{1+w} \quad , \tag{5.1}$$

the expressions (3.6) and (3.14) for the deceleration parameter and the statefinder parameter $r$ take the form

$$q = \frac{1}{2} \cdot (1 + 3w - 3\kappa\varsigma / H) ,  \tag{5.2}$$

$$r = 1 + \frac{9}{2} w(1+w) - \frac{9}{4}\left(1+3w - \frac{\kappa\varsigma}{H}\right)\frac{\kappa\varsigma}{H} . \tag{5.3}$$

The expression (5.3) may be factorized as

$$r = 1 + \frac{9}{4}\left(1+w - \frac{\kappa\varsigma}{H}\right)\left(2w - \frac{\kappa\varsigma}{H}\right) . \tag{5.4}$$

For such universe models the statefinder parameter $s$ is

$$s = \frac{1}{2} \cdot \frac{(1+w - \kappa\varsigma/H)(2w - \kappa\varsigma/H)}{w - \kappa\varsigma/H} . \tag{5.5}$$

The equation of motion of the cosmic expansion for the present class of models may be written

$$\dot{H} + \frac{3}{2}(1+w)H^2 - \frac{3}{2}\kappa\varsigma H = 0 . \tag{5.6}$$

For later comparison Brevik and Gorbunova first considered a universe model dominated by a perfect fluid with no viscosity, and wrote the solution of eq.(4.4) for that case as[23]

$$a(t) = a_0 \left[1 + \frac{3}{2} H_0 (1+w)(t - t_0)\right]^{-2/3\alpha} . \tag{5.7}$$

In this paper I will consider Big Bang Universe models with $a(0) = 0$. This demands that

$$H_0 t_0 = \frac{2}{3(1+w)} . \tag{5.8}$$

Normalizing the scale factor to unity at the present time, we then obtain

$$a = \left(\frac{t}{t_0}\right)^{\frac{2}{3(1+w)}} . \tag{5.9}$$

The general solution of eq.(5.6) with viscosity was written by Brevik as[25]

$$a = a_0 \left[1 + \frac{3}{2}(1+w)H_0 t_\varsigma \left(e^{(t-t_0)/t_\varsigma} - 1\right)\right]^{\frac{2}{3(1+w)}} , \quad t_\varsigma = \left(\frac{3}{2}\kappa\varsigma\right)^{-1} . \tag{5.10}$$

Demanding again $a(0) = 0$ requires

$$H_0 t_\varsigma = \frac{2}{3(1+w)\left(1-e^{-t_0/t_\varsigma}\right)} \quad . \tag{5.11}$$

With the normalization $a(t_0)=1$ the scale factor is

$$a = \left(\frac{e^{t/t_\varsigma}-1}{e^{t_0/t_\varsigma}-1}\right)^{\frac{2}{3(1+w)}} \quad . \tag{5.12}$$

The corresponding Hubble parameter is

$$H = H_0 \frac{1-e^{-t_0/t_\varsigma}}{1-e^{-t/t_\varsigma}} \quad . \tag{5.13}$$

Inserting eq.(5.13) in eqs. (5.3) and (5.5), and using eq.(5.11) give the expressions (4.13) and (4.14) for the statefinder parameter with $t_A$ replaced by $t_\varsigma$.

We see that the expansion behaviour of this model, dominated by a single viscous fluid, is given by identical expressions to those of the model considered in section 4, dominated by a fluid with a non-linear equation of state. This is, however very natural. The viscous fluid has an effective pressure given by eq.(3.3). The first Friedmann equation still has the form (4.2). Hence the effective pressure is

$$\bar{p} = p - 3\kappa\varsigma H = w\rho - \varsigma\kappa^{3/2}\sqrt{3\rho} \quad , \tag{5.14}$$

and the equation of continuity is

$$\dot{\rho} + \sqrt{3\kappa}(1+w)\rho^{3/2} - 3\kappa\varsigma\rho = 0 \quad , \tag{5.15}$$

which has the same form as eq.(4.3).

The initial value of the Hubble parameter is $H(0)=\infty$. When $t\to\infty$ the Hubble-parameter approaches the value.

$$H_\infty = H_0\left(1-e^{-t_0/t_\varsigma}\right) = \frac{2}{3(1+w)t_\varsigma} = \frac{\kappa\varsigma}{1+w} \quad . \tag{5.16}$$

The time $t_\varsigma$ depends upon the strength of the viscosity. From the expression of $t_\varsigma$ given in eq.(5.10) follows that a small viscosity implies a large value of $t_\varsigma$. Brevik[25] has estimated the value of $\varsigma$. At $t=1000s$ after the Big Bang he finds that $t_\varsigma \approx 10^{21}\,y$. Hence, we have that $e^{t/t_\varsigma} \approx 1$ for the whole history of the universe up to now. Eq.(5.11) then gives to lowest order in $t/t_\varsigma$,

$$w \approx \frac{2}{3H_0 t_0} - 1 \approx -\frac{1}{3} \quad . \tag{5.17}$$

For these models the observationally favoured value of the statefinder parameters, $(r,s)=(1,0)$ takes place at the points of time $t_{41}$ and $t_{42}$ given by

$$H(t_{41}) = \frac{\kappa\varsigma}{1+w} \quad , \quad H(t_{42}) = \frac{\kappa\varsigma}{2w} . \tag{5.18}$$

For $w<0$ only $t_{41}$ corresponds to an expanding universe. Comparing with eq.(5.16) we see that $H(t_{41})=H_\infty$. Hence, the statefinder parameters of these universe models approach the favoured value in the infinite future.

## 6. Conclusion

Brevik and co-workers have investigated several classes of universe models with different types of dark energy. Two of these classes, one with dark energy having an inhomogeneous equation of state, and one with viscous dark energy, have been considered in this paper, and it has been demonstrated that they are closely related.

While Brevik and co-workers focused upon the late-time behaviour of these models with special emphasis upon the question whether they would enter a so-called Big Rip singularity, I have focused upon these universe models as Big Bang models and their behaviour up to the present time.

The statefinder parameters of the models have been calculated, and some restrictions upon the dark energy has been obtained by demanding that they should pass through an era where the values of the statefinder parameter are not too far from the values $(r,s)=(1,0)$ of the $\Lambda CDM$-model that are favoured by observations.